\documentclass[fleqn,12pt,twoside]{article}
\usepackage{espcrc1}

\usepackage{graphicx}
\usepackage[figuresright]{rotating}

\newcommand{\AmS}{{\protect\the\textfont2
  A\kern-.1667em\lower.5ex\hbox{M}\kern-.125emS}}

\title{Detailed Lattice-QCD Study for the Three-Quark Potential and Y-type Flux-Tube Formation}

\author{H. Suganuma\address[TITech]{Faculty of Science, Tokyo Institute of Technology, \\ 
                                    Ohokayama 2-12-1, Tokyo 152-8551, Japan}, 
        T.T. Takahashi\address{Yukawa Institute for Theoretical Physics, Kyoto University, \\ 
                               Kitashirakawa, Sakyo, Kyoto 606-8502, Japan}
        and
        H. Ichie\addressmark[TITech]}
       
\begin{document}

\maketitle

\begin{abstract}
We study the ground-state three-quark (3Q) potential $V_{\rm 3Q}^{\rm g.s.}$
and the excited-state 3Q potential $V_{\rm 3Q}^{\rm e.s.}$ using SU(3) lattice QCD.
From the accurate and thorough calculation for 
more than 300 different patterns of 3Q systems, the static ground-state 3Q potential 
$V_{\rm 3Q}^{\rm g.s.}$ is found to be well described 
by the Coulomb plus Y-type linear potential, i.e., Y-ansatz, within 1\%-level deviation.
With lattice QCD, we calculate also 
the excited-state potential in the 3Q system, and  
find the gluonic excitation energy, $V_{\rm 3Q}^{\rm e.s.}-V_{\rm 3Q}^{\rm g.s.}$,
to be about 1 GeV. 
This large gluonic-excitation energy would play an essential role to  
the success of the quark model for the low-lying hadrons 
in terms of the absence of the gluonic mode.
\end{abstract}

\section{The Static Three-Quark Potential and the Y-type Flux-Tube Formation}
The three-quark (3Q) potential is one of the most important fundamental quantities in the hadron physics, 
because it is directly responsible to the baryon properties and 
is the key quantity to clarify the quark confinement in baryons.

We perform the first systematic study of the ground-state 3Q potential $V_{\rm 3Q}^{\rm g.s.}$
and the excited-state 3Q potential $V_{\rm 3Q}^{\rm e.s.}$
using SU(3) lattice QCD~\cite{TMNS01,TSNM02,TS03,TSIMN03}.
For more than 300 different patterns of the spatially-fixed 3Q systems, 
we perform the thorough calculation 
for the ground-state potential $V_{\rm 3Q}^{\rm g.s.}$ 
in lattice QCD with $12^3\times 24$ at $\beta=5.7$ and
with $16^3\times 32$ at $\beta=5.8, 6.0$ at the quenched level.
For the accurate calculation, we construct the quasi-ground-state of the 3Q system 
using the smearing method~\cite{TMNS01,TSNM02,TS03}.

As an important conclusion, the static ground-state 3Q potential $V_{\rm 3Q}^{\rm g.s.}$
is found to be well described by the ``Coulomb plus Y-type linear potential", i.e., Y-ansatz,
\begin{equation}
V_{\rm 3Q}^{\rm g.s}=-A_{\rm 3Q}\sum_{i<j}\frac1{|{\bf r}_i-{\bf r}_j|}
+\sigma_{\rm 3Q} L_{\rm min}+C_{\rm 3Q},
\end{equation}
within the 1\%-level deviation~\cite{TMNS01,TSNM02}. 
Here, $L_{\rm min}$ denotes the minimal value of total flux-tube length, as shown in Fig.1.
Such a Y-type flux-tube profile is actually observed in recent lattice QCD~\cite{TSIMN03,IBSS03}, as shown in Fig.2.
From the comparison with the Q-$\rm\bar Q$ potential, 
$V_{\rm Q\bar Q}(r)=-A_{\rm Q\bar Q}\frac1r
+\sigma_{\rm Q\bar Q} r+C_{\rm Q\bar Q}$, 
we find the universality of the string tension as 
$\sigma_{\rm 3Q}\simeq\sigma_{\rm Q\bar Q}$ 
and the one-gluon-exchange result as $A_{\rm 3Q}\simeq\frac12 A_{\rm Q\bar Q}$~\cite{TMNS01,TSNM02}.

\newpage

\begin{figure}[hb]
\centering
\includegraphics[width=1.75in]{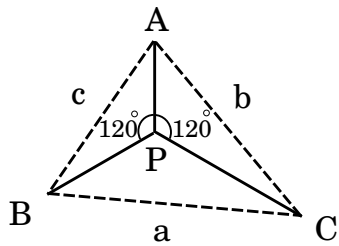}
\vspace{-1cm}
\caption{The flux-tube configuration of the 3Q system with the minimal value $L_{\rm min}$ of 
the total flux-tube length. There appears a physical junction linking 
the three flux tubes at the Fermat point P, and one finds $L_{\rm min}={\rm AP}+{\rm BP}+{\rm CP}$.
}
\label{fig1}
\end{figure}

\begin{figure}[hb]
\vspace{-1.5cm}
\centering
\includegraphics[width=4.5in]{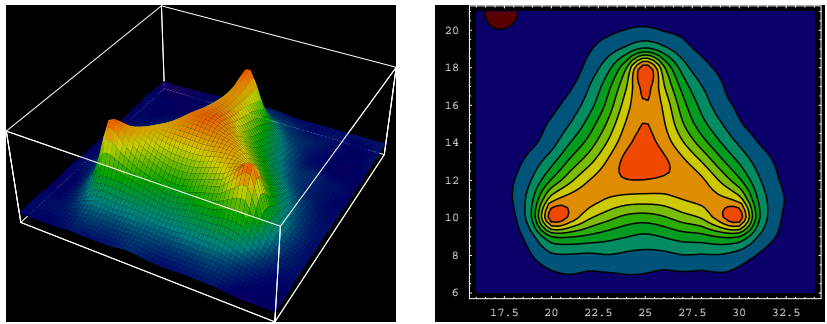}
\vspace{-0.8cm}
\caption{The lattice QCD result for the flux-tube profile 
in the spatially-fixed 3Q system with AP=BP=CP$\simeq$0.5fm 
in the maximally-abelian projected QCD~\cite{TSIMN03,IBSS03}.
}
\label{fig2}
\vspace{-1.3cm}
\end{figure}

\vspace{0.2cm}
\section{The Gluonic Excitation Energy and the Success of the Quark Model}

We study also the excited-state potential $V_{\rm 3Q}^{\rm e.s.}$ 
in the spatially-fixed 3Q systems in lattice QCD with $16^3\times 32$ at $\beta=5.8$~\cite{TS03,TSIMN03}.
The energy gap between $V_{\rm 3Q}^{\rm g.s.}$ and $V_{\rm 3Q}^{\rm e.s.}$ physically means 
the excitation energy of the gluon field in the presence of the spatially-fixed three quarks, 
and the gluonic excitation energy, $V_{\rm 3Q}^{\rm e.s.}-V_{\rm 3Q}^{\rm g.s.}$, is found to be about 1GeV, as shown in Fig.3.
Note that the gluonic excitation energy is rather large compared with the excitation energy of quark origin, 
and the present result predicts that the hybrid baryon expressed as $qqqG$ should be heavier than 2GeV.
%
%
%

\vspace{0.55cm}

\begin{figure}[hb]
\vspace{-1.3cm}
\centering
\includegraphics[width=2.4in]{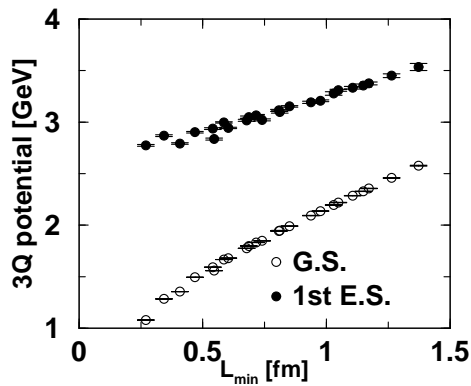}
\vspace{-1cm}
\caption{The lattice QCD results of 
the ground-state 3Q potential $V^{\rm g.s.}_{\rm 3Q}$ (open circles) and the 1st 
excited-state 3Q potential $V^{\rm e.s.}_{\rm 3Q}$ (filled circles) as the function of $L_{\rm min}$.
The gluonic excitation energy is found to be about 1GeV in the hadronic scale.
}
\label{fig3}
\end{figure}

\newpage

\begin{figure}[hb]
\vspace{-4cm}
\hspace{-1cm}
\includegraphics[width=7in]{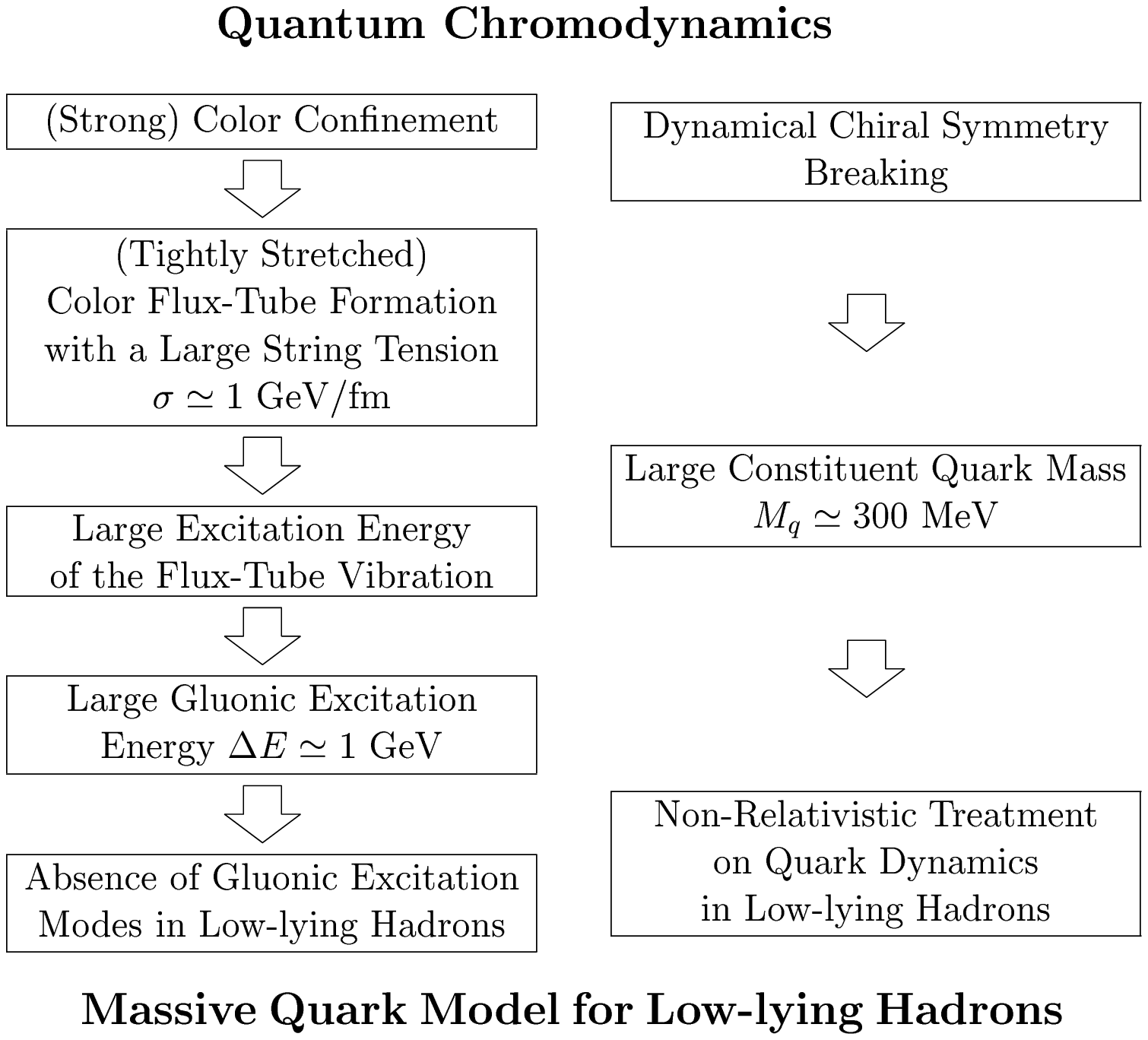}
\vspace{-11cm}
\caption{A possible scenario from QCD to the quark model in terms of 
color confinement and dynamical chiral-symmetry breaking (DCSB).
DCSB provides a large constituent quark mass of about 300MeV, which enables the non-relativistic treatment for quark dynamics. 
Color confinement provides the color flux-tube formation among quarks with a large string tension of $\sigma \simeq$ 1 GeV/fm.
In the flux-tube picture, the gluonic excitation is described as the flux-tube vibration, 
and the flux-tube vibrational energy is expected to be large, 
reflecting the large string tension.
The large gluonic-excitation energy of about 1GeV leads to 
the absence of the gluonic mode in low-lying hadrons, 
which would play the key role to the success of the quark model without gluonic excitation modes.
}
\vspace{-0.75cm}
\label{fig4}
\end{figure}


We show in Fig.4 a possible scenario from QCD to the massive quark model.
In terms of the flux-tube picture, the large gluonic-excitation energy is conjectured to originate from the large string tension 
as a result of strong color confinement.
The large gluonic-excitation energy would be responsible to the absence of the gluonic mode and 
the success of the quark model for low-lying hadrons.

\end{document}